%% file: acl_latex.tex
\pdfoutput=1

\documentclass[11pt]{article}

\usepackage[preprint]{acl}

\usepackage{times}
\usepackage{latexsym}

\usepackage[T1]{fontenc}

\usepackage[utf8]{inputenc}

\usepackage{microtype}

\usepackage{inconsolata}

\usepackage{graphicx}
\usepackage{multirow} 
\usepackage{booktabs} 
\usepackage{bbding}
\usepackage{amssymb}
\usepackage{colortbl}
\usepackage{amsmath} 
\usepackage[normalem]{ulem}
\usepackage{inconsolata}
\usepackage{arydshln}
\usepackage{pifont}
\useunder{\uline}{\ul}{}

\title{FEA-Bench: A Benchmark for Evaluating Repository-Level Code Generation for Feature Implementation}

\author{
Wei Li\textsuperscript{\rm $\heartsuit$}\thanks{Work done during internship at MSRA.}, 
Xin Zhang\textsuperscript{\ding{171}}\thanks{Corresponding author.}\thanks{Project leader.}, 
Zhongxin Guo\textsuperscript{\ding{171}},
Shaoguang Mao\textsuperscript{\ding{171}} \\
\textbf{Wen Luo}\textsuperscript{\rm $\heartsuit$}\textbf{,}
\textbf{Guangyue Peng}\textsuperscript{\rm $\heartsuit$}\textbf{,}
\textbf{Yangyu Huang}\textsuperscript{\ding{171}}\textbf{,}
\textbf{Houfeng Wang}\textsuperscript{\rm $\heartsuit$ \textdagger}\textbf{,}
\textbf{Scarlett Li}\textsuperscript{\ding{171}} \\
{\textsuperscript{\rm $\heartsuit$} State Key Laboratory of Multimedia Information Processing,} \\
{School of Computer Science,  Peking University} \\ 
{\textsuperscript{\ding{171}} Microsoft Research Asia} \\
{\tt weili22@stu.pku.edu.cn } \\
{\tt xinzhang3@microsoft.com, wanghf@pku.edu.cn }
}

\begin{document}
\maketitle
\begin{abstract}

Implementing new features in repository-level codebases is a crucial application of code generation models. However, current benchmarks lack a dedicated evaluation framework for this capability. To fill this gap, we introduce FEA-Bench, a benchmark designed to assess the ability of large language models (LLMs) to perform incremental development within code repositories. We collect pull requests from 83 GitHub repositories and use rule-based and intent-based filtering to construct task instances focused on new feature development. Each task instance containing code changes is paired with relevant unit test files to ensure that the solution can be verified. 
The feature implementation requires LLMs to simultaneously possess code completion capabilities for new components and code editing abilities for other relevant parts in the code repository, providing a more comprehensive evaluation method of LLMs' automated software engineering capabilities.
Experimental results show that LLMs perform significantly worse in the FEA-Bench, highlighting considerable challenges in such repository-level incremental code development. 
Our code will soon be publicly available at \href{https://github.com/microsoft/FEA-Bench}{https://github.com/microsoft/FEA-Bench}.

\end{abstract}

\section{Introduction}

The remarkable text generation capabilities of large language models (LLMs) \cite{achiam2023gpt4report} have extended their impact into the domain of code generation \cite{xu2022codellms_review}, which has led to the emergence of developer assistants such as Copilot, Cursor, Devin, etc. An important research topic is evaluating the effectiveness of LLMs in generating code across diverse scenarios. Many of the existing benchmarks focus on evaluating standalone programming problems, such as HumanEval, MBPP, and LiveCodeBench \cite{chen2021humaneval, austin2021mbpp, jain2024livecodebench}. These benchmarks offer little insight into the challenges developers face in real-world projects, where codebases are composed of multiple interconnected files. In such projects, modifications in one part of the code often necessitate corresponding edits elsewhere. This type of collaborative, large-scale development is referred to as repository-level code development.

\begin{figure}[t]
    \centering
    \includegraphics[width=1.0\columnwidth]{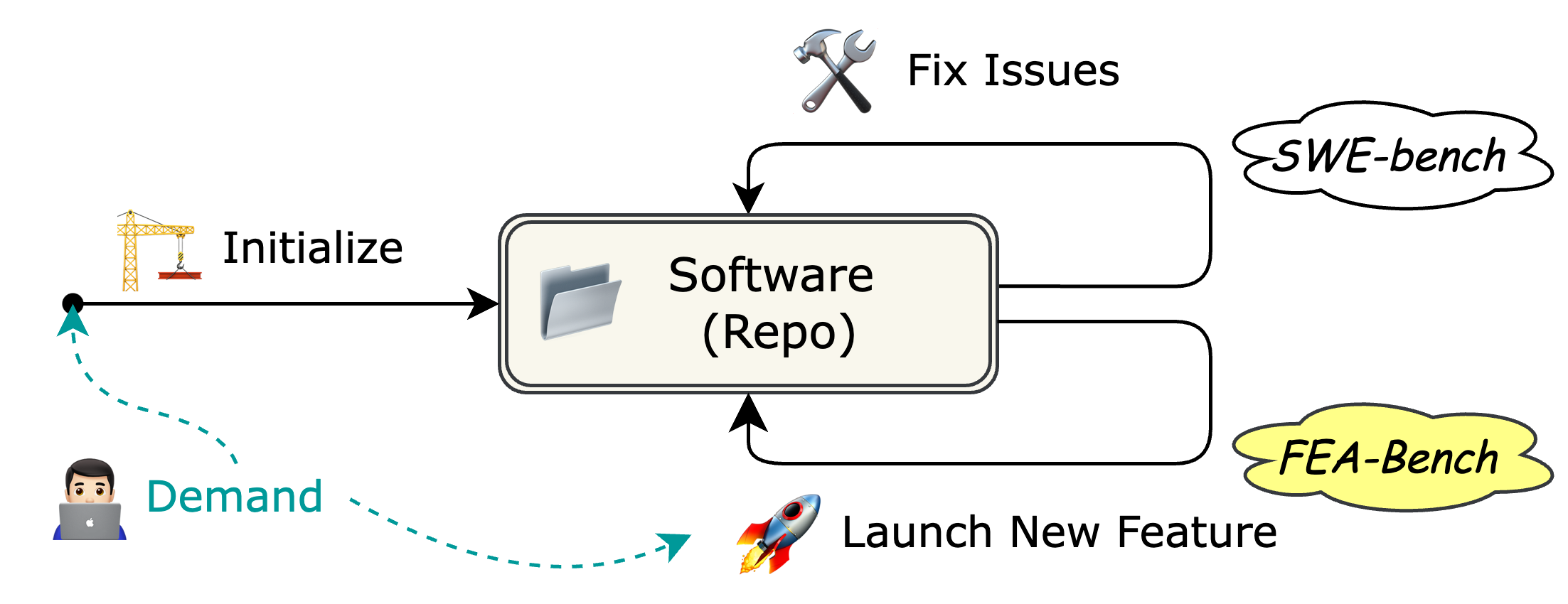}
    \caption{The proposed FEA-Bench aims to evaluate incremental repository development, while SWE-bench \cite{jimenez2024swebench} focuses on repairing issues.}
    \label{fig:intro}
\end{figure}

In the realm of repository-level code generation, much of the current evaluation effort is centered around code completion \cite{li2024deveval, yang2024execrepobench}. Code completion refers to generating correct code snippets at specified locations within a given code context. However, this task is inherently limited—it typically targets localized generation and does not account for broader implications beyond the scope of completion. 

Recent advancements in the capabilities of LLMs have expanded their potential role from merely suggesting code snippets to managing the full lifecycle of repository development. A prominent benchmark in this domain is SWE-bench \cite{jimenez2024swebench}, which evaluates LLMs on resolving issues, primarily focusing on bug fixes within repositories. In practice, as shown in Figure \ref{fig:intro}, a more critical aspect of software engineering is the launch of new features, which often entails introducing new functions or even entire files into the repository. The continuous implementation of new features drives software growth and is a key focus of automated software engineering. We define such tasks as repository-level incremental code development. In this work, to bridge the gap in benchmarks for this domain, we construct a dataset derived from pull requests in GitHub repositories that specifically focus on adding new components, with the overarching goal of implementing new features. Each task instance in our dataset is paired with its corresponding unit test files, culminating in the creation of the \textbf{Fea}ture Implementation \textbf{Bench}mark (\textbf{FEA-Bench}), which comprises 1,401 task instances sourced from 83 diverse GitHub repositories.

Statistically, our dataset exhibits characteristics that significantly differ from the bug-fix-oriented SWE-bench. Task instances in FEA-Bench require the implementation of new functions and classes in Python and involve substantially longer code generation compared to SWE-Bench. Experimental results demonstrate that current LLMs perform poorly on the proposed benchmark. According to our execution-based metrics, the best-performing LLM, DeepSeek-R1, successfully resolves only about 10\% of the task instances. The key contributions of this paper are as follows:

\begin{itemize}
\item We introduce the task of repository-level incremental code development, addressing a critical challenge in real-world software engineering where the continuous implementation of new features is essential for sustaining software growth.
\item We construct the first benchmark to evaluate repository-level incremental code development. By employing parsing and other filtering methods, we constructed a dataset composed of feature implementation tasks, offering execution-based evaluation.
\item Using an automated pipeline, we scale the test data to include 83 diverse code repositories, ensuring high diversity. We will publicly release our data collection and evaluation codebase, allowing FEA-Bench to be continuously updated and expanded.
\end{itemize}

\section{Related Work}

\subsection{Code Large Language Models}
Large language models (LLMs) have revolutionized software engineering by enabling code generation, debugging, and translation capabilities \cite{pan2024codev-bench, li2023codeeditor, joshi2023coderepair, shi2024hierarchical-debugging}. Large-scale pre-trained LLMs such as GPT-4 \cite{achiam2023gpt4report}, CodeLlama \cite{roziere2023codellama}, DeepSeek-Coder \cite{guo2024deepseek-coder, zhu2024deepseekcoder-v2}, and Qwen2.5-Coder \cite{hui2024qwen2.5-coder} have demonstrated proficiency in generating functional code across multiple programming languages. Recent advancements also include instruction-tuned models like Starcoder \cite{li2023starcoder}, WizardCoder \cite{luo2023wizardcoder}, WaveCoder \cite{yu-etal-2024-wavecoder}, Magicoder \cite{wei2024magicoder} and EpiCoder \cite{wang2025epicoder}. With these advancements, code LLMs are poised to further revolutionize how developers interact with code, promising increased efficiency in software creation. Additionally, the integration of agents \cite{luo2025agentreview} has further enhanced the performance of LLMs in software engineering tasks \cite{yang2024swe-agent, xia2024agentless, AutoCodeRover}. In this paper, we further evaluate the performance of current LLMs in the incremental code development scenarios at the repository level. This investigation aims to drive the research on code LLMs toward addressing more intricate software engineering challenges, thereby advancing the capabilities of these models in handling sophisticated development tasks.

\subsection{Code Generation Benchmarks}
Recent code generation benchmarks, such as HumanEval \cite{chen2021humaneval} and MBPP \cite{austin2021mbpp}, have primarily focused on synthesizing standalone functions or scripts from natural language, while subsequent efforts like APPS \cite{hendrycks2021apps}, EvalPlus \cite{liu2024evalplus}, CoderEval \cite{yu2024codereval},  ClassEval \cite{du2023classeval}, BigCodeBench \cite{zhuo2024bigcodebench}, and FullStackBench \cite{liu2024fullstackbench} have expanded evaluation to more complex scenarios. However, these benchmarks largely overlook the repository-level challenges of real-world software development, a gap addressed by recent works \cite{bairi2024codeplan, zhang2023repocoder}. Because of the wild applications of auto code completion tools like Github Copilot \cite{dakhel2023github-copilot}, most repository-level code generation benchmarks aim to evaluate the code completion capabilities of LLMs \cite{liu2023repobench}.  Repository-level code completion aims to generate code for incomplete code snippets within a repository \cite{wang2024rlcoder}. Benchmarks such as DevEval \cite{li2024deveval}, EvoCodeBench \cite{li2024evocodebench}, Codev-Bench \cite{pan2024codev-bench}, and ExecRepoBench \cite{yang2024execrepobench} evaluate this capability. However, code completion data constructed by removing a single line or function body can suffer from future context leakage issues \cite{zheng2024humanevo}.

In practice, autonomous development of a code repository should include all edits to the code, rather than simply completing the specified code under perfect context. SWE-bench \cite{jimenez2024swebench} focuses on repairing repositories' issues by revising existing programs. This highlights the need for models to effectively modify and integrate changes within the repository. Our work fills another gap of incremental repository-level development by introducing a benchmark that evaluates LLMs on the implementation of new features, further bridging the divide between code completion and real-world software engineering.

\section{Benchmark Construction}

\subsection{Overview}
\label{sec:data-overview}
\begin{figure}[t]
    \centering
    \includegraphics[width=1.0\columnwidth]{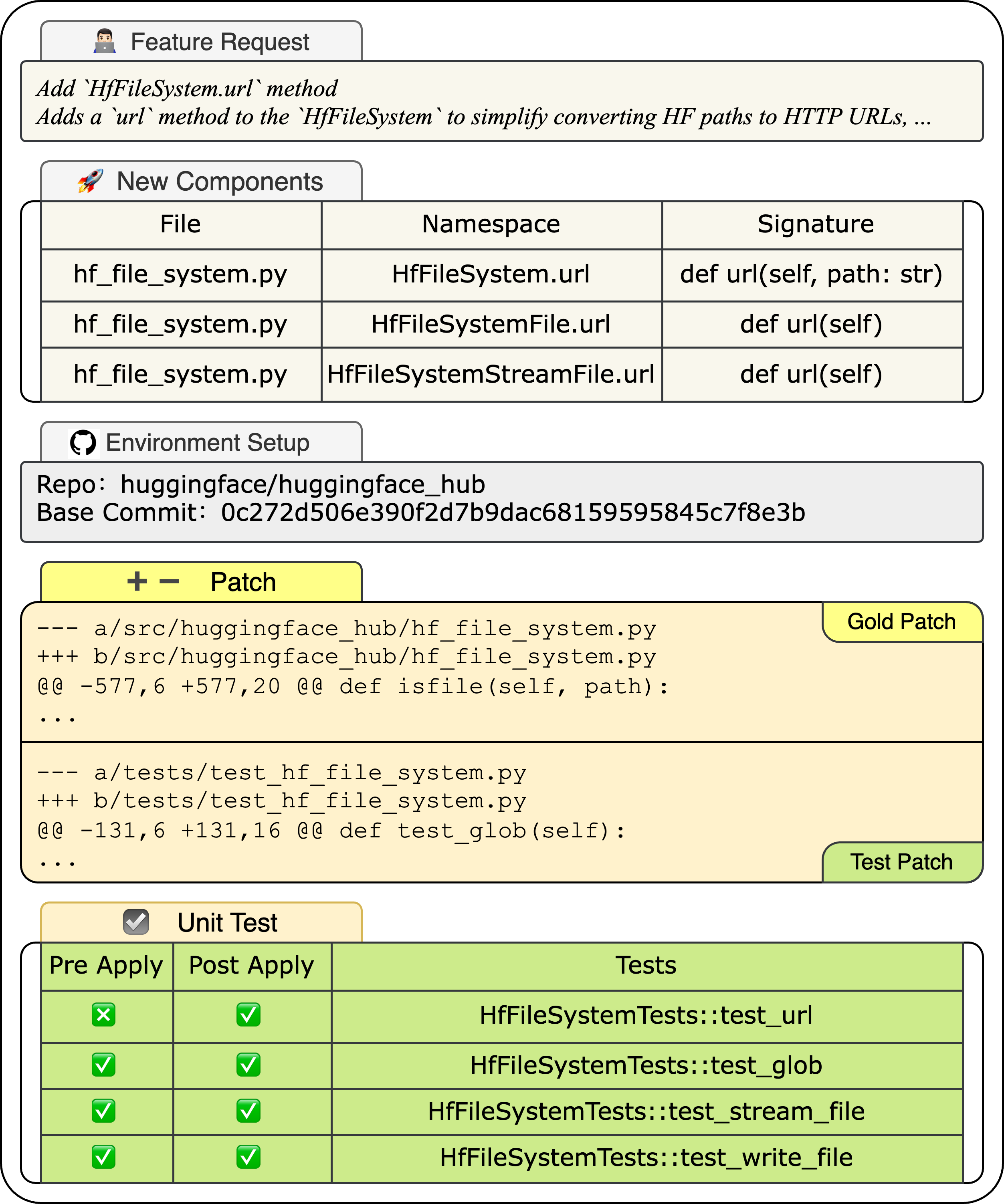}
    \caption{An example of the task instances from the FEA-Bench. During the inference of LLMs, the first two items: feature request and new components are considered as known information. The environment setup serves as a prerequisite for creating the testbed and environment. Python file patches and unit tests are used as labels and evaluation metrics and should not be leaked during the inference of LLMs.
}
    \label{fig:instance}
\end{figure}

The task instances in FEA-Bench are constructed based on existing pull request (PR) data from GitHub. As illustrated in Figure \ref{fig:instance}, each task instance contains the following elements:

\textbf{\ding{182} Feature Request} Content of the pull request and corresponding issues (if any) provide essential information regarding the new feature or functionality to be developed.

\textbf{\ding{183} Definition of New Components} This includes the signatures and documentation of newly added functions and classes. The name of the new component must be consistent with that in PR, which is the prerequisite for completing the task, because unit tests are written based on the specified name.

\textbf{\ding{184} Environment Setup} This includes the relevant information for the repository and specifies the base commit of the code repository for each task instance. Besides, the configurations of the building execution environment are also included.

\textbf{\ding{185} Patch} It describes the changes made to the code in the repository and can be processed by the \texttt{unidiff} standard library\footnote{\href{https://github.com/matiasb/python-unidiff}{https://github.com/matiasb/python-unidiff}}. Additionally, the changes can be applied to the repository using the \texttt{git apply} tool. A patch can be divided into a test patch and a gold patch; the former pertains to changes in test codes, while the latter involves the other changes affecting the software itself.

\textbf{\ding{186} Unit Test} The correctness of the code changes is verified based on the result of running these tests. We get the ground truth status by actually running \texttt{pytest} before and after applying the gold patch.

\begin{figure*}[ht]
    \centering
    \includegraphics[width=1.0\textwidth]{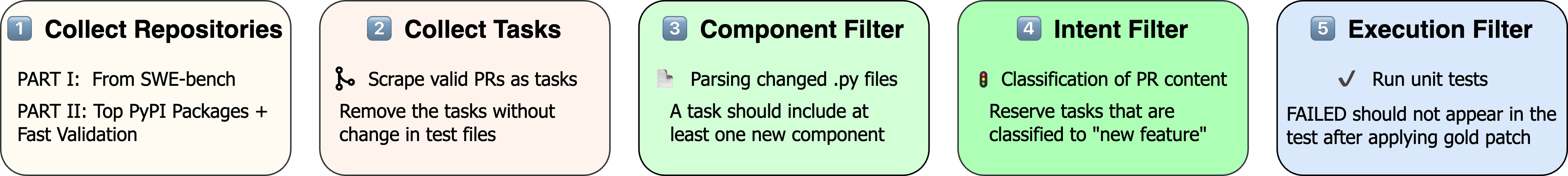}
    \caption{The data collection pipeline for the FEA-Bench. First, determine the scope of GitHub repositories from which task instances will be collected. Next, gather pull requests as task instances and apply filtering criteria to select instances that meet the purpose of adding new features. Finally, use the included test files to execute unit tests, ensuring that only task instances with reproducible test results are included in FEA-Bench. For a more detailed construction process, please refer to Appendix \ref{sec:appendix-dataset-construction}.
}
    \label{fig:collection}
\end{figure*}

Our data collection pipeline is developed based on SWE-bench. As shown in Figure \ref{fig:collection}, the proposed collection pipeline ensures data diversity and comprehensiveness, enabling a robust evaluation of LLMs' capabilities in implementing new features at the repository level. By making our data collection and evaluation codes publicly available, we aim to facilitate continuous updates and the creation of new versions of FEA-Bench. In the rest of this section, we will discuss the construction and the characteristics of FEA-Bench.

\subsection{Repository Collection}
\label{sec:repo-collection}
To determine the scope of GitHub repositories for data collection, we initially focus on the GitHub repositories corresponding to Python packages listed on the Top PyPI website\footnote{\href{https://hugovk.github.io/top-pypi-packages/}{https://hugovk.github.io/top-pypi-packages/}}. Packages appearing on this list are generally influential Python software with high data quality, and most repositories use a unified \text{pytest} format for testing, which facilitates later execution. This leaderboard contains 8,000 Python packages. We obtain approximately 600 repositories that meet the criteria of having a license and more than 1,000 pull requests by applying a filtering process.

\textbf{Fast Validation} Except for repositories already included in SWE-bench, the remaining repositories do not have customized installation procedures or testing methods. While environment configurations vary among different Python packages, \texttt{pip} offers a unified installation approach\footnote{\texttt{pip install -e .}}. And we adopt \texttt{pytest} as the default method for unit testing. Based on these settings, for each repository, we extract the first 20 pull requests that include changes to test files and observe the unit test status. Repositories that have at least one task instance where the unit tests passed with the default configuration are retained, as they possess the potential to generate task instances that meet our criteria. 

After this fast validation process, in addition to the 18 Python packages included in the SWE-bench dataset, another 101 Python packages are identified as sources for extracting repository-level data for our benchmark.

\subsection{Task Collection and Filtering}
\label{sec:task-collection}
Based on existing repository data, we crawl all pull requests and consider those that include changes to test files as possible task instances. Given our focus on the incremental feature development task, we introduce the following steps to obtain the final task instances for FEA-Bench through filtering:

\textbf{Extraction of new components} For each task instance candidate, we perform parsing on all Python scripts involved in the gold patch. We compare the state before and after applying the patch to identify newly added components, including classes and functions, and extract their signatures and docstrings as metadata.

\textbf{Filtering based on new components} We retain only those task instances that contain at least one new component. To ensure that implementing new features is the primary purpose of the pull request, we further restrict the new components to occupy more than 25\% of all edited lines in gold patch. This threshold is set relatively low because code changes often need to include modifications to other related code in addition to the new components themselves.

\textbf{Intent-based filtering} The pull requests filtered using the above rule-based approach still include some that are not primarily aimed at feature implementation. Therefore, we use GPT-4o to classify the intent based on the pull request description. Only pull requests classified as "new feature" are retained.

\textbf{Verification by running unit tests} For each instance, we first set up the environment and testbed. We then apply the test patch and run the unit test files involved in the test patch. Theoretically, some unit tests should fail at this stage. After applying the gold patch, we rerun the same unit tests. If the configuration is correct, all unit tests should pass. Unlike SWE-bench, we do not impose restrictions on whether \texttt{ImportError} or \texttt{AttributeError} occurs before applying the gold patch, as these errors are almost inevitable before new components are implemented. Instead, to ensure data quality, we exclude samples where tests remain in the failed status after applying the gold patch. The status of each test function before and after applying the gold patch is recorded.

After task collection and filtering, 83 out of 119 repositories collected by the method in Section \ref{sec:repo-collection} produce 1,401 task instances for new feature implementation.

\input{table/stats-compare}

\subsection{Benchmark Characteristics}

\textbf{Semi-guided software engineering task} Although the signatures for new components are provided, the primary objective of the FEA-Bench is to evaluate the whole solution of new feature implementation. This is a comprehensive real-world software engineering task, distinct from code completion. As shown in Table \ref{tab:stat-compare}, each instance in FEA-Bench involves an average of 128.5 modified lines, with 87.1 lines attributed to the new components themselves. This indicates that approximately 41.4 lines of changes are made elsewhere in the repository. Implementing new features not only requires the ability to generate code for specified new components but also necessitates making complementary changes within the existing repository.

\textbf{Lite version} We have also curated a subset to serve as a lite version of our dataset. This subset is filtered based on criteria including higher quality and lower difficulty. This lite version is particularly useful for evaluating systems that are computationally intensive and time-consuming. Detailed information can be found in Section \ref{sec:appendix-lite}.

\textbf{New components driven generation task}
From Table \ref{tab:stat-compare}, we can observe that, on average, the number of lines for new components in each FEA-Bench task instance is more than $ 8\times $ that of SWE-bench. Furthermore, the new components account for approximately 67.8\% of all edited lines in FEA-Bench, compared to just 28.9\% in SWE-bench. The difference in this metric between FEA-Bench lite and SWE-bench verified is even more pronounced. These statistics indicate that the task instances in FEA-Bench are primarily aimed at implementing new features. In SWE-bench, the average number of new functions is 0.73, indicating that its task instances mainly involve editing existing code rather than incremental development. 

\textbf{Complex solutions} While SWE-bench focuses on fixing issues, which generally involve simpler problems, the task instances in FEA-Bench exhibit greater complexity. From Table \ref{tab:stat-compare}, whether measured by the number of edited lines or edited files, the solutions of task instances in FEA-Bench are notably more complex than those in SWE-bench.

\section{Experimental Design}

\subsection{Models}

Due to its repository-level code generation characteristics, FEA-Bench requires models to have a long context window. We evaluate representative code LLMs and general-purpose LLMs with strong foundational capabilities on the FEA-Bench. The code LLMs used in our evaluation include CodeLlama, Codestral\footnote{\href{https://mistral.ai/en/news/codestral}{https://mistral.ai/en/news/codestral}}, Qwen2.5-Coder, and DeepSeek-Coder-V2 \cite{roziere2023codellama, hui2024qwen2.5-coder, zhu2024deepseekcoder-v2}. For general-purpose LLMs, we evaluate the performance of GPT-4, GPT-4o \cite{achiam2023gpt4report, GPT-4o} and DeepSeek-V3 \cite{liu2024deepseek-v3}, as well as models with long chain-of-thought (CoT) capabilities such as o1 and DeepSeek-R1 \cite{jaech2024openai-o1, guo2025deepseek-r1}.

\subsection{Context}
\label{sec:setting-context}
\input{table/main_results}

To explore the capabilities and potential limits of LLMs in implementing new features within code repositories, we construct different prompts from several perspectives based on our collected data. Each task instance is evaluated using various context settings to provide a comprehensive understanding of model performance.

\textbf{New component hints} In FEA-Bench, information about new components can be derived from two sources: 1) signatures and documentation of newly extracted functions and classes, and 2) changes in non-Python files within the patch, which often contain relevant information about the new components. Based on this, we have two settings: \textbf{Brief.} Only provides the signatures of new components. \textbf{Detailed.} Includes all the aforementioned information.

\textbf{Retrieval method} Given the extensive amount of code across multiple files in a repository, selective inclusion of file contents as context is necessary. Firstly, the README file and files containing new components are always included in the context. For other files, similar to SWE-bench, we divide retrieval methods into: \textbf{Oracle.} Includes all files involved in the patch in the context. \textbf{BM25.} Retrieves relevant files across the entire repository by BM25 algorithm \cite{robertson2009bm25} based on the content of the pull request, and ranks them by relevance and filling the context until reaching a specified length.

\textbf{Output format} Generating an entire file can lead to interruptions due to generation limits and is costly. Therefore, our experiments offer two edit-based generation settings: \textbf{Natural.} Generates code edits in a natural format as pairs of before-and-after snippets, which can be converted into patches applicable to the code repository through post-processing. \textbf{Patch.} Directly generates edits in patch format. Since patches use line numbers for fragment location, this setting includes line numbers in the context's code content.

The details of prompt and experimental settings are shown in Appendix \ref{sec:appendix-inference}.

\section{Evaluation Results}

\textbf{The performance of LLMs} The evaluation results of FEA-Bench are presented in Table \ref{tab:main_results}. It is observed that in \textit{Oracle} and \textit{Detailed} prompt settings, the best resolved ratio of task instances is 9.92\%, indicating the poor performance of LLMs in the incremental development task at the repository level. Generally, the models with larger parameter sizes demonstrate better results. Among code LLMs, Qwen2.5-Coder exhibited performance comparable to that of GPT-4,  highlighting its superiority in the domain of code generation. Despite this, general-purpose LLMs with stronger foundational capabilities can approach the performance of specialized code LLMs. For example, R1-Distill, which shares the same underlying architecture as Qwen2.5-Coder, showed competitive performance. Among the models evaluated, the latest DeepSeek-V3 and R1 models achieve the best performance, significantly outperforming OpenAI's GPT-4 and o1 series. This underscores the importance of foundational capabilities in LLMs for repository-level development tasks. Additionally, the performance on the lite version is slightly higher, but the relative trends remain largely consistent. When computational resources are limited, the metrics from FEA-Bench lite can be used to reflect the performance on the full benchmark.

\textbf{The performance under different contexts} 
As shown in Table \ref{tab:main_results}, we evaluate the performance of LLMs under different settings of new component hints and retrieval methods. Overall, detailed new component hints lead to better model performance. However, in the results from FEA-Bench lite, brief hints that only provide signatures performed better. This discrepancy could be attributed to the lack of structured presentation of new components and their documentation within the prompt, as illustrated in Figure \ref{fig:prompt}. Regarding retrieval methods, the \textit{Oracle} setting generally outperforms the \textit{BM25} setting, although the difference is not substantial. This may be because code files containing new components, which are regarded as known conditions in the FEA-Bench, ensure a certain baseline performance. Considering the simplest instances in the dataset: those involving modifications to only one code file. Since instances in FEA-Bench always include new components, the unique code file is known information. In this case, \textit{BM25} retrieved files supplement additional information, which can lead to better model performance compared to the \textit{Oracle} setting.

\section{Discussion}
\input{table/retrieval_analysis}
\input{table/output_format}
\input{table/agent_results}
\begin{figure*}[ht]
    \centering
    \includegraphics[width=1.0\textwidth]{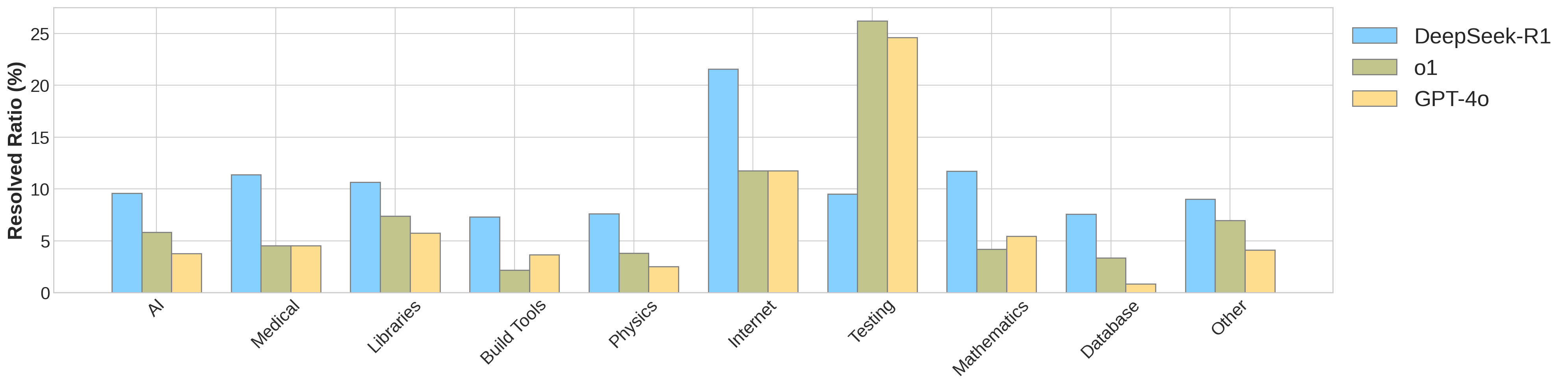}
    \caption{The resolved ratios grouped by the categories of the repositories.
}
    \label{fig:result-class}
\end{figure*}

\subsection{Retrieved Files in Context}
We aim to investigate whether providing more code context helps LLMs improve new feature development. Therefore, we conducted experiments by increasing the prompt limit from 27K to 40K tokens, which surpasses the original context window limit of Qwen2.5-Coder but can contain more retrieved files. The results are presented in Table \ref{tab:retrieval-analysis}. The evaluated models include GPT-4 and GPT-4o, and we also report the average precision and recall of Python files in the context relative to those in the gold patch, as well as the proportion of instances where all involved files are recalled. 

Although the recalls slightly improve at the limit of 40K tokens, model performance decreases. This indicates that current LLMs still struggle to extract useful information from long contexts in repository-level code development tasks. To improve performance on FEA-Bench, enhancing the precision of retrieval may be an effective approach than simply increasing the context length.

\subsection{Output Format of Edits}
\label{sec:output_format}
Direct repository-level code development outputs edits rather than new code itself. How LLMs can better generate edits remains an open question. Therefore, we analyze the impact of output formats using the two configurations described in Section \ref{sec:setting-context}. The two rightmost columns of Table \ref{tab:output-format} illustrate a comparison between the \textit{Natural} and the \textit{Patch} generation method, with the former demonstrating significantly higher performance. The possible reason is that generating patches imposes stricter formatting requirements, which current LLMs find challenging to adhere to accurately. Therefore, the main results presented in Table \ref{tab:main_results} adopt the performance by \textit{Natural} generation method.

Whether using patches converted by converting before-and-after code snippets in \textit{Natural} mode or directly generating patches in \textit{Patch} mode, during evaluation, these patches must be applied to the repository. The success rates of the \texttt{git apply} are shown in the middle two columns of Table \ref{tab:output-format}. It can be observed that the success rate in the \textit{Natural} prompt mode is significantly higher, contributing to the superior performance of LLMs in \textit{Natural} mode. Further observation reveals a significant positive correlation between the success rate of applying patches and the resolved ratio, regardless of whether \textit{Natural} or \textit{Patch} mode is used. DeepSeek-R1, which performs the best on FEA-Bench, has the highest values for both metrics. This indicates that the format of code edits is a critical factor limiting the performance of LLMs on such tasks.

\subsection{Evaluation of Agent Frameworks}
To further evaluate the performance of current state-of-the-art methods on FEA-Bench, we conduct experiments on the lite subset using the cost-efficient Agentless and Agentless-lite frameworks \cite{xia2024agentless}. We compare the final performance of using these agents with that of directly constructing the context using \textit{Oracle} and \textit{BM25} retrieval methods, as shown in Table~\ref{tab:agent_results}. Notably, \textit{Agentless} improves the resolved ratio of task instances over \textit{BM25} for most models, with the largest gains observed for GPT-4o, o1-mini, and o1. This improvement strongly correlates with the increased success rate of applying code edits, highlighting the challenge of adhering to the required code editing format, as discussed in Section~\ref{sec:output_format}. Overall, Agentless and Agentless-lite perform similarly, with no significant advantage over the results under the \textit{Oracle} setting. This indicates that current methods still have substantial room for improvement in repository-level new feature implementation.

\subsection{Performances across Repositories}
\label{sec:performance-class}
To further analyze LLMs' performance at a finer granularity, we examine resolved ratios across different categories of repositories. We classify repositories into several categories, as shown in Appendix \ref{sec:appendix-dataset-repositories}. The performance of DeepSeek-R1, o1, and GPT-4o across different categories is illustrated in Figure \ref{fig:result-class}. The resolved ratios of different models vary across categories; task instances in the \textit{Testing} category (from the repo: \textit{joke2k/faker}) have the highest resolved ratio, followed by the \textit{Internet} category. For the remaining categories, the pass ratio of the three models are at a similar level.

Among the three LLMs, GPT-4o shows slightly weaker performance compared to the other models, but its trend is largely consistent with o1. Notably, DeepSeek-R1 exhibits weaker performance in the \textit{Testing} category but significantly outperforms both GPT-4o and o1 in all other categories. This suggests that integrating different models might further enhance overall performance on FEA-Bench.

\subsection{Performances under Different Complexity of New Components}

\begin{figure}[t]
    \centering
    \includegraphics[width=1.0\columnwidth]{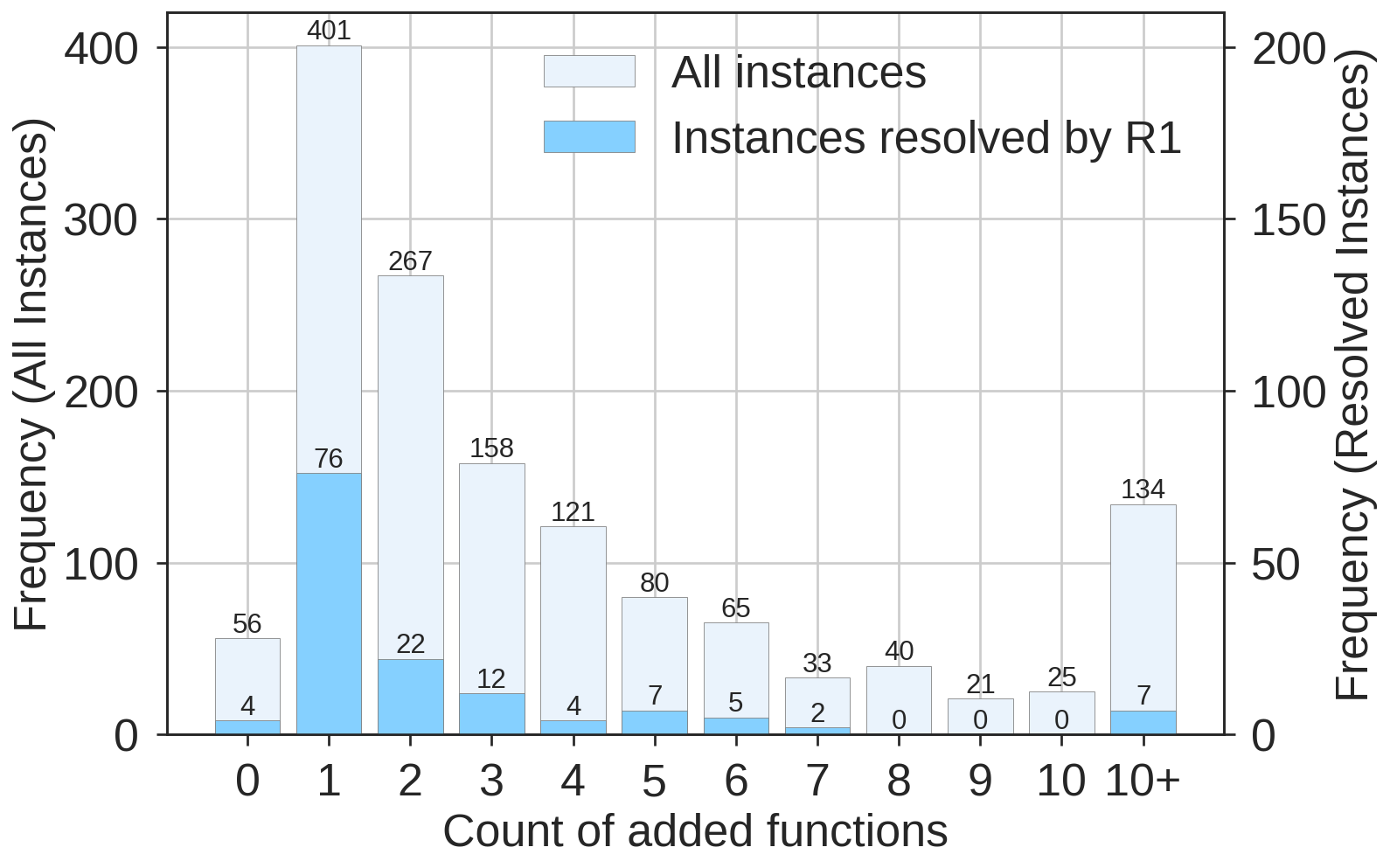}
    \caption{Histogram of the number of added functions both in all task instances and resolved task instances by DeepSeek-R1 (under \textit{Natural} and \textit{Detailed} prompt settings).}
    \label{fig:added-components}
\end{figure}
To further investigate whether the implementation of new components adds pressure on LLMs in completing feature implementation tasks, we examine the relationship between the number of new functions and the number of resolved instances. The number of added functions is more indicative of the complexity of new components, as functions are more atomic compared to classes. Figure \ref{fig:added-components} illustrates the distribution of task instances and those solved by R1 with respect to the number of added functions. Excluding instances where the number of added functions is zero (i.e., adding only classes for storing variables), it is evident that the resolved ratio decreases as the number of added functions increases. The resolved ratio is 18.96\% when the number of added functions is 1, 8.24\% when it is 2, and 5.47\% when it is greater than or equal to 3. This indicates that implementing new features is a more challenging task than fixing bugs, which constitute the majority of instances in SWE-bench. Moreover, the more complex the new components, the higher the difficulty of successful implementation.

\section{Conclusion}
In this paper, we introduce FEA-Bench, a novel benchmark for evaluating the repository-level incremental code development capabilities of large language models (LLMs). Our benchmark focuses on the critical task of implementing new features by adding new components to existing code repositories. Through our comprehensive dataset and rigorous evaluation, we demonstrate that current LLMs face significant challenges in this domain. We also analyzed that the retrieval method of files, the output format, the repository itself, and the complexity of new components all impact the implementation of new features. Our work highlights the need for further advancements in LLMs' reasoning and generation capabilities to better address real-world software engineering tasks. We hope that FEA-Bench will serve as a valuable resource for the research community, driving progress in this important area.

\section*{Limitations}
Our constructed data and experiments have certain limitations. First, our benchmark includes only Python repositories, as Python projects are easier to execute and generally follow consistent testing frameworks. Second, the quantity of high-quality data suitable for repository-level incremental development is limited. High-quality and usable pull requests for new feature development are relatively scarce. Many repository-level code developments for implementing new functionalities were committed during the early stages of repositories, without going through the rigorous code review process typical of the open-source community, resulting in lower data quality that cannot be utilized. Furthermore, the software's early-stage developments might not even have been conducted using the GitHub platform, posing a challenge for data collection and utilization. Consequently, FEA-Bench, which is built on publicly available data and subjected to stringent filtering, may exhibit certain scenario limitations.

Due to the long context involved in repository-level code development, the cost of conducting experiments using LLMs is relatively high. Therefore, the experimental results are based on a single round generation, akin to Pass@1, which may introduce a certain level of bias into the results. Additionally, given the scarcity of API resources for models like DeepSeek-V3 and R1, some results in the main experiments presented in Table \ref{tab:main_results} are missing. We hope that more affordable models similar to DeepSeek can be further developed to facilitate research and applications in the field of repository-level code development.

\section*{Ethics Statement}
We collected the data from publicly available Github repositories only for research purposes. All the repositories have licenses that allow free software use. LLMs are used only for classification during the construction of the FEA-Bench dataset, so no harmful information can be created in the dataset.  The dataset and code for our proposed method will be made publicly available for academic research. However, we should note that the inference results of the task instances from the benchmark may contain code that is harmful to computer systems. Evaluation by docker is recommended, just like SWE-bench.

Additionally, the ChatGPT platform was used as an AI assistant for refining the paper writing.

\section*{Acknowledgments}
This work was supported by National Science and Technology Major Project (No. 2022ZD0116308)  and National Natural Science Foundation of China (62036001) . The corresponding author is Xin Zhang and Houfeng Wang.

\bibliography{custom}

\appendix

\section{Dataset Details}

\begin{figure*}[ht]
    \centering
    \includegraphics[width=0.9\textwidth]{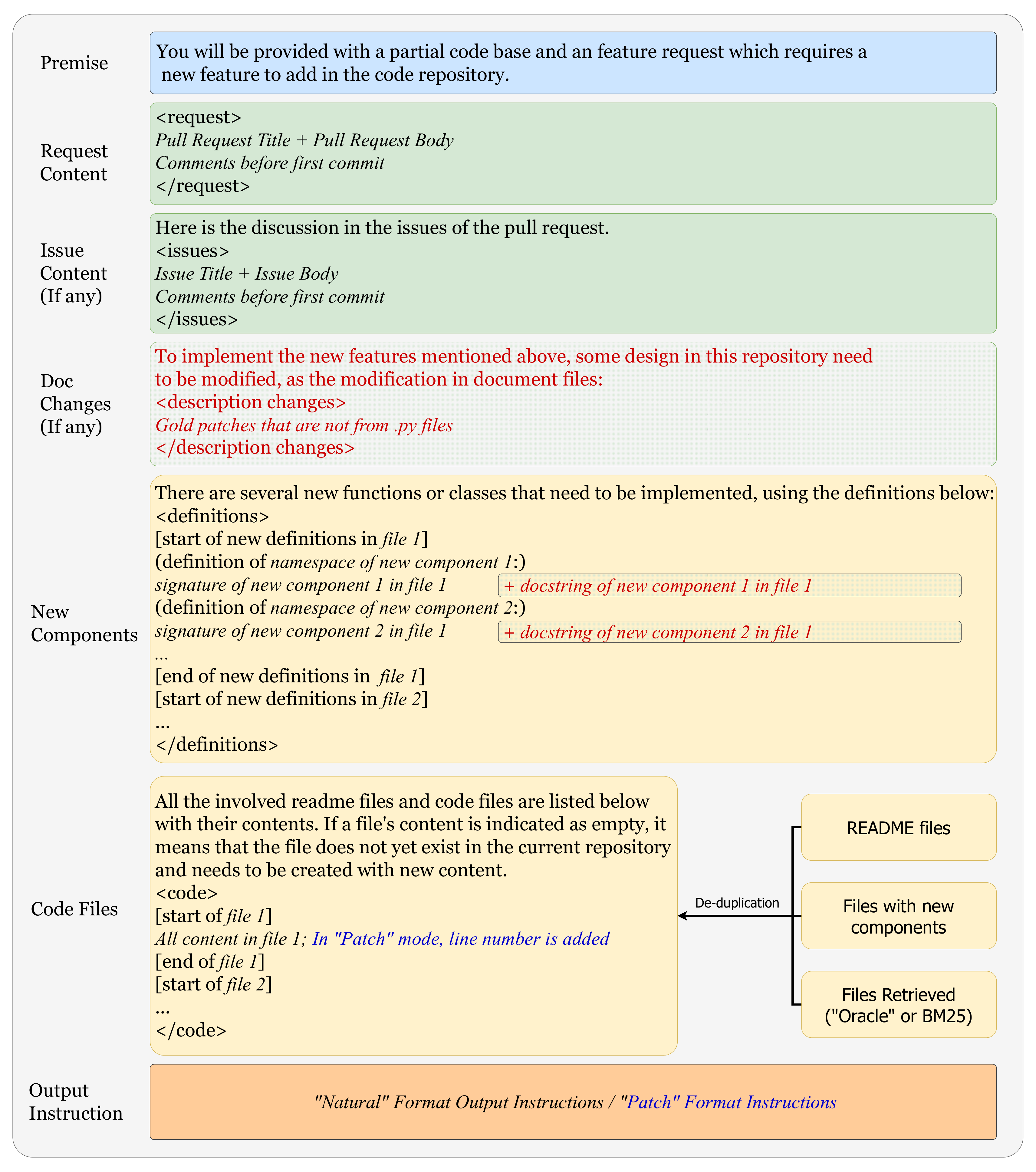}
    \caption{The prompt for the inference of the task instances.
}
    \label{fig:prompt}
\end{figure*}

\begin{figure*}[ht]
    \centering
    \includegraphics[width=1.0\textwidth]{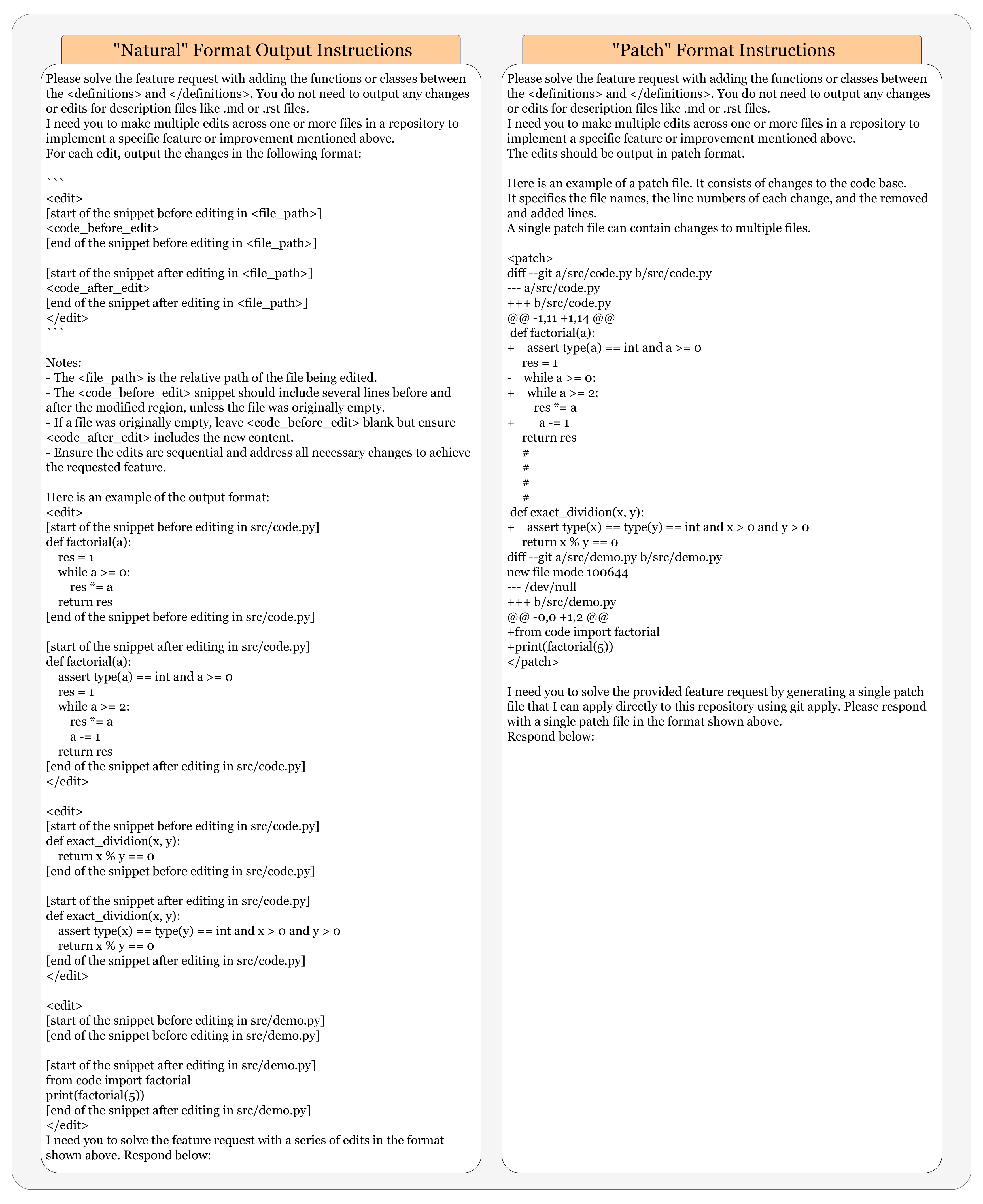}
    \caption{The output instructions at the rear of the inference prompt.
}
    \label{fig:output-instruction}
\end{figure*}

\subsection{Repositories}
\label{sec:appendix-dataset-repositories}
\input{table/repo_info}
The task instances in FEA-Bench are derived from 83 Python packages corresponding to GitHub repositories. For each package, we obtained its license information and topic classification from the PyPI website. For packages missing the topic attribute, we utilized their README files and employed GPT-4o to classify them into topics available on the PyPI website\footnote{\href{https://pypi.org/classifiers/}{https://pypi.org/classifiers/}}. Finally, considering the distribution of the topics of task instances in FEA-Bench, we further simplify the topics to several categories to facilitate data visualization and analysis, as presented in Section \ref{sec:performance-class}.

The relevant information regarding the involved code repositories is summarized in Table \ref{tab:repo-info}.

\subsection{Construction Details}
\label{sec:appendix-dataset-construction}
\input{table/repo_filter_stat}

Table \ref{tab:repo-filter-stat} shows the number of remaining task instances for each stage during the data collection process. In Section \ref{sec:repo-collection}, we mentioned that after the fast validation step, including the 18 repositories from SWE-bench, a total of 119 repositories (18 + 101) are available for further data collection. On this basis, we crawl the pull requests from GitHub to obtain code changes. Files with names containing words like \textit{test} are identified as unit test files executable by \texttt{pytest}.

Initially, we filter pull requests (PRs) based on whether they were merged, resulting in the number of PRs shown in the "\# PR" column of Table \ref{tab:repo-filter-stat}. Next, we excluded PRs without any test files, leaving only those PRs with the necessary conditions to be considered valid task instances, as indicated by the "\# All tasks" column. Further filtering was conducted according to the third and fourth stages illustrated in Figure \ref{fig:collection}, retaining only those PRs that introduce at least one new component and are classified as "new feature" types by GPT-4o based on the PR content. Additionally, instances with patch lengths exceeding 8K (8192) tokens are excluded to remove long-tail distributions and noise. The remaining candidate task instances are listed under the "\# Candidates" column. This filtering step does not exist in SWE-bench, leading to a smaller number of candidate task instances in our benchmark dataset.

Finally, for each instance, we apply the test patch to the repository in the base commit state to verify that unit tests could accurately evaluate the corresponding code edits. Similar to SWE-bench, our pipeline annotates each instance with the corresponding repository version and provides environment and installation configurations based on versions. For repositories included in SWE-bench, we directly utilize their environment setup and test configurations grouped by versions. For other repositories, minimal installation instructions (\texttt{pip install -e .}) and basic \texttt{pytest} testing configurations are used. Initially, we run the unit tests directly and observe the pass status. Subsequently, we apply the gold patch from the PR and execute the unit tests again. We allowed errors such as \texttt{AttributeError} and \texttt{ImportError} in the first test, which are common when new components are not yet implemented, but these errors are not permitted in SWE-bench data collection pipeline. For the second test after applying the gold patch, any FAILED status is unacceptable to ensure data quality. After this execution-based filtering process, the final task instances constitute the FEA-Bench task instances, as shown in the "\# in Full" column of Table \ref{tab:repo-filter-stat}.

We only present information for repositories that have at least one instance in FEA-Bench. Specifically, out of the 119 repositories identified during the repository collection phase, only 83 repositories contain at least one task instance.

\subsection{FEA-Bench lite}
\label{sec:appendix-lite}

The feature implementation task proposed in this study, as one primary type of repository-level incremental code development tasks, require LLMs with long context capabilities to perform reasoning over extensive file contents. Such inference is computationally expensive. Considering the possible evaluation of multi-round code generation systems, it is necessary to select a high-quality subset for more efficient evaluation. Therefore, we establish stricter criteria to curate a higher-quality, lower-difficulty FEA-Bench lite subset.

Instances meeting any of the following low-quality criteria are excluded:
\begin{itemize}
    \item The feature request descriptions contain fewer than 40 words.
    \item The instance involves cascading issues or commit SHA-256 references.
    \item The descriptions contain images.
\end{itemize}

Additionally, to limit task difficulty, instances meeting any of the following criteria are also excluded:
\begin{itemize}
    \item Involve deleting code files.
    \item Involve more than three code files.
    \item The gold patch contains More than 10 code change hunks.
    \item Natural-formatted code change content exceeding 4K(4096) tokens.
    \item Contain new class(es).
    \item Contain more than ten added functions.
\end{itemize}

Beyond the lite subset, we also expect to collaborate with professional software engineers to annotate a verified subset, similar to SWE-bench Verified\footnote{\href{https://openai.com/index/introducing-swe-bench-verified/}{https://openai.com/index/introducing-swe-bench-verified/}}.

\section{Inference}
\label{sec:appendix-inference}

\subsection{Prompt}
\label{sec:appendix-prompt}
In Section \ref{sec:setting-context}, we present different prompt settings for the context of inference. A more detailed prompt structure is illustrated in Figure \ref{fig:prompt}. The prompt is constructed using the first two items shown in Figure \ref{fig:instance} as known input information. The feature request includes both pull request content and issue content. Information about new components and related code files is listed straightforwardly within the prompt.

In Figure \ref{fig:prompt}, the italicized text indicates placeholders that need to be filled with specific data of each task instance, while the other text represents standard prompt content. Red text highlights additional information for detailed hints of new components, and blue text indicates parts that need to be modified when directly generating patches as results.

In this study, we provide two formats for generating code edits: \textit{Natural} and \textit{Patch}. The corresponding output instructions are illustrated in Figure \ref{fig:output-instruction}. To ensure that the models produce outputs in the correct format, both modes are accompanied by detailed instructions and examples.

We include several different prompts mentioned above for each task instance in the dataset files.

\subsection{Generation Configurations}

For models with fewer than 32 billion parameters, we utilized the vLLM framework \footnote{\href{https://github.com/vllm-project/vllm}{https://github.com/vllm-project/vllm}} on an 8-GPU NVIDIA A100 workstation, employing tensor parallelism for inference. The maximum number of generated tokens is limited to 4096, matching the generation length limits of used OpenAI GPT-4 and GPT-4o in our experiments.

For larger open-source and all closed-source models, specifically DeepSeek-V3, R1, and OpenAI series models, we invoke their APIs for inference. The versions of the OpenAI models used are as follows:
\begin{itemize}
    \item GPT-4: \textit{gpt-4-turbo-2024-04-09}
    \item GPT-4o: \textit{gpt-4o-2024-05-13}
    \item o1: \textit{o1-2024-12-17}
    \item o1-mini: \textit{o1-mini-2024-09-12}
\end{itemize}

When possible, the temperature and top-p settings are fixed at 0.2 and 0.95, respectively. For DeepSeek-V3 and R1, the max output tokens are 8K (8192). For o1 and o1-mini, the max output tokens is 100,000 and 65,536 (64K), respectively.

During the inference process, LLMs perform a single generation for each task instance in FEA-Bench, and the output is converted into a patch for evaluation. The evaluation tools are adapted from the SWE-bench evaluation scripts which are based on docker, ensuring safety and easy use for the evaluation process.

\end{document}

%% file: table/stats-compare.tex
\begin{table}
\centering
\resizebox{\columnwidth}{!}{
\begin{tabular}{lcccc}
\toprule
 & \multicolumn{2}{c}{\textbf{FEA-Bench}} & \multicolumn{2}{c}{\textbf{SWE-bench}} \\ \cline{2-5} 
\textbf{} & \textbf{Full} & \textbf{Lite} & \textbf{Full} & \textbf{Verified} \\ \midrule
$ \left| \mathrm{Repositories} \right| $  & 83 & 48 & 12 & 12 \\
$ \left| \mathrm{Tasks} \right| $  & 1401 & 200 & 2294 & 500 \\ \midrule
\# Lines of oracle files & 2115.5 & 1366.8 & 1961.3 & 1488.2 \\
\# Files edited & 2.62 & 1.54 & 1.66 & 1.24 \\
\# Lines edited & 128.5 & 68.1 & 37.71 & 14.32 \\ \midrule
\# Lines of added components & 87.1 & 47.2 & 10.9 & 2.1 \\
\% Added components & 67.8 & 69.3 & 28.9 & 14.7 \\
\# Functions added & 4.49 & 2.02 & 0.73 & 0.25 \\
\# Classes added & 0.78 & 0 & 0.064 & 0.006 \\ \bottomrule
\end{tabular}
}
\caption{
Statistics for FEA-Bench and its lite subset, as well as for SWE-bench and its verified subset. The metrics include: 1) the number of task instances and involved repositories; 2) the average total number of lines in all Python files involved in code changes (i.e., the \textit{Oracle} setting described in Section \ref{sec:setting-context}), and the average number of edited files and lines; 3) the average number of lines of new components, the average percentage of new component lines relative to all edited lines, and the average number of added functions and classes.
}
\label{tab:stat-compare}
\end{table}

%% file: table/main_results.tex
\begin{table*}[ht]
\centering
\resizebox{\textwidth}{!}{
\begin{tabular}{lllcccccccc}
\toprule
 &  &  & \multicolumn{4}{c}{\textbf{FEA-Bench}} & \multicolumn{4}{c}{\textbf{FEA-Bench lite}} \\ \cline{4-11} 
\textbf{Model} & \textbf{Size} & \textbf{Window} & \multicolumn{2}{c}{\textbf{Oracle}} & \multicolumn{2}{c}{\textbf{BM25 (27K)}} & \multicolumn{2}{c}{\textbf{Oracle}} & \multicolumn{2}{c}{\textbf{BM25 (27K)}} \\ \cline{4-11} 
 &  &  & \textbf{Detailed} & \textbf{Brief} & \textbf{Detailed} & \textbf{Brief} & \textbf{Detailed} & \textbf{Brief} & \textbf{Detailed} & \textbf{Brief} \\ \midrule
\multirow{2}{*}{\textbf{CodeLlama}} & 13B & 16K & 0.14 & 0.43 & $\times$ & $\times$ & 0.0 & 0.0 & $\times$ & $\times$ \\
 & 34B & 16K & 0.57 & 0.57 & $\times$ & $\times$ & 0.0 & 0.0 & $\times$ & $\times$ \\ \midrule
\multirow{2}{*}{\textbf{Qwen2.5-Coder}} & 14B & 32K & 3.57 & 3.57 & 3.71 & 2.93 & 3.5 & 4.5 & 3.5 & 2.5 \\
 & 32B & 32K & 4.43 & 3.64 & 3.85 & 2.78 & 6.0 & 5.5 & 2.0 & 2.5 \\ \midrule
\textbf{Codestral-22B} & 22B & 32K & 0.86 & 0.93 & 1.43 & 1.36 & 0.5 & 1.0 & 0.0 & 0.0 \\ \midrule
\textbf{DeepSeek-Coder-V2} & 16B & 128K & 0.21 & 0.29 & 0.57 & 0.36 & 0.0 & 0.5 & 0.0 & 0.5 \\
\textbf{DeepSeek-R1-Distill} & 32B & 128K & 3.78 & 4.07 & 4.78 & 4.21 & 5.5 & 7.5 & 7.0 & 5.5 \\
\textbf{DeepSeek-V3} & 671B & 64K & 8.14 & 6.92 & 8.21 & 7.64 & 14.5 & 10.5 & \textbf{13.0} & 12.0 \\
\textbf{DeepSeek-R1} & 671B & 64K & \textbf{9.92} & \textbf{8.35} & \textbf{10.49} & \textbf{9.85} & \textbf{14.5} & \textbf{14.5} & 12.0 & \textbf{13.5} \\ \midrule
\textbf{GPT-4} &  & 128K & 4.71 & 4.21 & 3.14 & 2.86 & 6.0 & 6.5 & 2.0 & 1.0 \\
\textbf{GPT-4o} &  & 128K & 6.14 & 5.57 & 5.28 & 4.50 & 5.0 & 5.0 & 4.0 & 3.5 \\
\textbf{o1-mini} &  & 128K & 1.93 & 1.86 & 2.28 & 2.57 & 2.0 & 3.5 & 1.0 & 2.0 \\
\textbf{o1} &  & 200K & 7.28 & 6.57 & 6.78 & 6.64 & 10.0 & 12.5 & 5.0 & 7.0 \\ \bottomrule
\end{tabular}
}
\caption{The resolved ratios on FEA-Bench (lite) task instances. The evaluation is conducted on single-round generation outputs by each model and a task instance is considered resolved only if all unit tests are passed. The prompt using BM25 retrieved files is limited to length of 27K tokens. This ensures that, with a maximum generation of 4K tokens, the total length will not exceed 32K tokens, which is the context window limits of most tested models. "Detailed" and "Brief" refer to the levels of hints regarding new components in prompt, as mentioned in Section \ref{sec:setting-context}.
}
\label{tab:main_results}
\end{table*}

%% file: table/retrieval_analysis.tex
\begin{table}
\centering
\resizebox{\columnwidth}{!}{
\begin{tabular}{lcccc}
\toprule
 & \multicolumn{2}{c}{\textbf{BM25 (27K)}} & \multicolumn{2}{c}{\textbf{BM25 (40K)}} \\ \cline{2-5} 
 & \multicolumn{4}{c}{\textbf{Retrieval Metrics (\%)}} \\ \hline
\textbf{Precision Avg.} & \multicolumn{2}{c}{40.26} & \multicolumn{2}{c}{31.85} \\
\textbf{Recall Avg.} & \multicolumn{2}{c}{76.04} & \multicolumn{2}{c}{77.14} \\
\textbf{Recall All} & \multicolumn{2}{c}{51.61} & \multicolumn{2}{c}{53.03} \\ \midrule
\textbf{} & \multicolumn{4}{c}{\textbf{Resolved Ratio (\%)}} \\
\textbf{} & \textbf{Detailed} & \textbf{Brief} & \textbf{Detailed} & \textbf{Brief} \\ \midrule
\textbf{GPT-4} & \textbf{3.14} & \textbf{2.86} & 3.14 & 2.78 \\
\textbf{GPT-4o} & \textbf{5.28} & 4.50 & 4.78 & \textbf{4.64} \\ \bottomrule
\end{tabular}
}
\caption{
The retrieval metrics and the instance resolved ratios under 27K and 40K token length limits of the prompt in \textit{BM25} retrieval mode.
}
\label{tab:retrieval-analysis}
\end{table}

%% file: table/output_format.tex
\begin{table}
\centering
\resizebox{\columnwidth}{!}{
\begin{tabular}{lcccc}
\toprule
 & \multicolumn{2}{c}{\textbf{\%Apply}} & \multicolumn{2}{c}{\textbf{\%Resolved}} \\ \cline{2-5} 
\textbf{Model} & \textbf{Natural} & \textbf{Patch} & \textbf{Natural} & \textbf{Patch} \\ \midrule
\textbf{Qwen2.5-Coder(32B)} & 44.82 & 12.92 & 4.43 & 1.71 \\
\textbf{R1-Distill(32B)} & 55.75 & 19.06 & 3.78 & 1.71 \\
\textbf{GPT-4} & 59.10 & \textbf{33.26} & 4.71 & \textbf{3.07} \\
\textbf{GPT-4o} & 66.38 & 19.49 & 6.14 & 1.86 \\
\textbf{o1} & 57.03 & - & 7.28 & - \\
\textbf{DeepSeek-V3} & 69.09 & - & 8.14 & - \\
\textbf{DeepSeek-R1} & \textbf{73.16} & - & \textbf{9.92} & - \\ \bottomrule
\end{tabular}
}
\caption{
The impact of output formats \textit{Natural} and \textit{Patch}. We show the success rates of applying code edits to the code repository and the final resolved ratios. The experiments on directly generating patches are conducted only on the models shown in the first four columns.
}
\label{tab:output-format}
\end{table}

%% file: table/agent_results.tex
\begin{table*}[ht]
\centering
\small
\resizebox{0.8\textwidth}{!}{
\begin{tabular}{lcccccccc}
\toprule
 & \multicolumn{2}{c}{\textbf{Agentless}} & \multicolumn{2}{c}{\textbf{Agentless-Lite}} & \multicolumn{2}{c}{\textbf{Oracle}} & \multicolumn{2}{c}{\textbf{BM25 (27K)}} \\ 
 & \textbf{\%Reso.} & \textbf{\%Apply} & \textbf{\%Reso.} & \textbf{\%Apply} & \textbf{\%Reso.} & \textbf{\%Apply} & \textbf{\%Reso.} & \textbf{\%Apply} \\ \midrule
\textbf{DeepSeek-V3} & - & - & \textbf{11.0} & 71.0 & \textbf{14.5} & 70.5 & \textbf{13.0} & 69.0 \\
\textbf{GPT-4} & - & - & 3.5 & 94.5 & 6.0 & 54.0 & 2.0 & 37.0 \\
\textbf{GPT-4o} & 9.0 & 87.5 & 9.5 & 96.0 & 5.0 & 60.0 & 4.0 & 53.5 \\
\textbf{o1-mini} & 4.5 & 69.5 & 4.5 & 81.5 & 2.0 & 32.5 & 1.0 & 23.5 \\
\textbf{o1} & \textbf{14.0} & 90.5 & 10.0 & 89.5 & 10.0 & 49.5 & 5.0 & 33.5 \\ \bottomrule
\end{tabular}
}
\caption{
The performance of Agentless on FEA-Bench lite, compared to the direct retrieval by \textit{Oracle} and \textit{BM25}. "\%Reso." refers to the resolved ratio of task instances, while "\%Apply" indicates the success rate of applying the generated code edits to the repository.
}
\label{tab:agent_results}
\end{table*}

%% file: table/repo_info.tex
\begin{table*}[ht]
\centering
\resizebox{0.9\textwidth}{!}{
\begin{tabular}{lllll}
\hline
\multicolumn{1}{c}{\textbf{Repo Name}} & \multicolumn{1}{c}{\textbf{License}} & \multicolumn{1}{c}{\textbf{Topic}} & \multicolumn{1}{c}{\textbf{Category}} & \multicolumn{1}{c}{\textbf{Source}} \\ \hline
astropy/astropy & BSD-3-Clause & Scientific/Engineering::Astronomy & Physics & SWE-Bench \\
django/django & BSD-3-Clause & Internet::WWW/HTTP & Internet & SWE-Bench \\
matplotlib/matplotlib & Other & Scientific/Engineering::Visualization & Other & SWE-Bench \\
mwaskom/seaborn & BSD-3-Clause & Scientific/Engineering::Visualization & Other & SWE-Bench \\
pallets/flask & BSD-3-Clause & Internet::WWW/HTTP & Internet & SWE-Bench \\
pvlib/pvlib-python & BSD-3-Clause & Scientific/Engineering::Physics & Physics & SWE-Bench \\
pydata/xarray & Apache-2.0 & Scientific/Engineering::Information Analysis & Other & SWE-Bench \\
pydicom/pydicom & Others & Scientific/Engineering::Medical Science Apps. & Medical & SWE-Bench \\
pylint-dev/astroid & LGPL-2.1 & Software Development::Libraries & Libraries & SWE-Bench \\
pylint-dev/pylint & GPL-2.0 & Software Development::Quality Assurance & Other & SWE-Bench \\
pyvista/pyvista & MIT & Scientific/Engineering::Information Analysis & Other & SWE-Bench \\
scikit-learn/scikit-learn & BSD-3-Clause & Scientific/Engineering::Artificial Intelligence & AI & SWE-Bench \\
sphinx-doc/sphinx & BSD-2-Clause & Text Processing::Markup & Other & SWE-Bench \\
sqlfluff/sqlfluff & MIT & Software Development::Quality Assurance & Other & SWE-Bench \\
sympy/sympy & Others & Scientific/Engineering::Mathematics & Mathematics & SWE-Bench \\ \hline
Aider-AI/aider & Apache-2.0 & Software Development::Code Generators & Other & Fast-Validation \\
Cog-Creators/Red-DiscordBot & GPL-3.0 & Communications::Chat & Other & Fast-Validation \\
DLR-RM/stable-baselines3 & MIT & Scientific/Engineering::Artificial Intelligence & AI & Fast-Validation \\
EleutherAI/lm-evaluation-harness & MIT & Scientific/Engineering::Artificial Intelligence & AI & Fast-Validation \\
Project-MONAI/MONAI & Apache-2.0 & Scientific/Engineering::Medical Science Apps. & Medical & Fast-Validation \\
PyThaiNLP/pythainlp & Apache-2.0 & Text Processing::Linguistic & Other & Fast-Validation \\
RDFLib/rdflib & BSD-3-Clause & Software Development::Libraries & Libraries & Fast-Validation \\
Textualize/rich & MIT & Software Development::Libraries & Libraries & Fast-Validation \\
Textualize/textual & MIT & Software Development::User Interfaces & Other & Fast-Validation \\
TileDB-Inc/TileDB-Py & MIT & Software Development::Libraries & Libraries & Fast-Validation \\
astronomer/astronomer-cosmos & Apache-2.0 & Software Development::Build Tools & Build Tools & Fast-Validation \\
atlassian-api/atlassian-python-api & Apache-2.0 & Internet::WWW/HTTP & Internet & Fast-Validation \\
aws-cloudformation/cfn-lint & MIT-0 & Software Development::Quality Assurance & Other & Fast-Validation \\
aws-powertools/powertools-lambda-python & MIT-0 & Software Development::Libraries & Libraries & Fast-Validation \\
aws/sagemaker-python-sdk & Apache-2.0 & Scientific/Engineering::Artificial Intelligence & AI & Fast-Validation \\
biopragmatics/bioregistry & MIT & Scientific/Engineering::Bio-Informatics & Other & Fast-Validation \\
boto/boto3 & Apache-2.0 & Software Development::Libraries & Libraries & Fast-Validation \\
boto/botocore & Apache-2.0 & Software Development::Libraries & Libraries & Fast-Validation \\
cocotb/cocotb & BSD-3-Clause & Scientific/Engineering::Electronic Design Automation (EDA) & Other & Fast-Validation \\
conan-io/conan & MIT & Software Development::Build Tools & Build Tools & Fast-Validation \\
deepset-ai/haystack & Apache-2.0 & Scientific/Engineering::Artificial Intelligence & AI & Fast-Validation \\
docker/docker-py & Apache-2.0 & Software Development::Libraries & Libraries & Fast-Validation \\
dpkp/kafka-python & Apache-2.0 & Software Development::Libraries & Libraries & Fast-Validation \\
embeddings-benchmark/mteb & Apache-2.0 & Scientific/Engineering::Artificial Intelligence & AI & Fast-Validation \\
facebookresearch/hydra & MIT & Software Development::Libraries & Libraries & Fast-Validation \\
fairlearn/fairlearn & MIT & Scientific/Engineering::Artificial Intelligence & AI & Fast-Validation \\
falconry/falcon & Apache-2.0 & Internet::WWW/HTTP & Internet & Fast-Validation \\
google-deepmind/optax & Apache-2.0 & Scientific/Engineering::Artificial Intelligence & AI & Fast-Validation \\
googleapis/python-aiplatform & Apache-2.0 & Scientific/Engineering::Artificial Intelligence & AI & Fast-Validation \\
googleapis/python-bigquery & Apache-2.0 & Internet::WWW/HTTP & Internet & Fast-Validation \\
gradio-app/gradio & Apache-2.0 & Scientific/Engineering::Human Machine Interfaces & Other & Fast-Validation \\
graphql-python/graphene & MIT & Software Development::Libraries & Libraries & Fast-Validation \\
huggingface/accelerate & Apache-2.0 & Scientific/Engineering::Artificial Intelligence & AI & Fast-Validation \\
huggingface/datasets & Apache-2.0 & Scientific/Engineering::Artificial Intelligence & AI & Fast-Validation \\
huggingface/huggingface\_hub & Apache-2.0 & Scientific/Engineering::Artificial Intelligence & AI & Fast-Validation \\
huggingface/pytorch-image-models & Apache-2.0 & Software Development::Libraries & Libraries & Fast-Validation \\
huggingface/trl & Apache-2.0 & Scientific/Engineering::Artificial Intelligence & AI & Fast-Validation \\
joblib/joblib & BSD-3-Clause & Software Development::Libraries & Libraries & Fast-Validation \\
joke2k/faker & MIT & Software Development::Testing & Testing & Fast-Validation \\
lark-parser/lark & MIT & Text Processing::Linguistic & Other & Fast-Validation \\
minio/minio-py & Apache-2.0 & Software Development::Libraries & Libraries & Fast-Validation \\
open-mmlab/mmengine & Apache-2.0 & Utilities & Other & Fast-Validation \\
openvinotoolkit/datumaro & MIT & Scientific/Engineering::Image Processing & Other & Fast-Validation \\
pgmpy/pgmpy & MIT & Scientific/Engineering::Artificial Intelligence & AI & Fast-Validation \\
pre-commit/pre-commit & MIT & Software Development::Quality Assurance & Other & Fast-Validation \\
prometheus/client\_python & Apache-2.0 & System::Monitoring & Other & Fast-Validation \\
prompt-toolkit/python-prompt-toolkit & BSD-3-Clause & Software Development::User Interfaces & Other & Fast-Validation \\
pygments/pygments & BSD-2-Clause & Software Development::Documentation & Other & Fast-Validation \\
pyocd/pyOCD & Apache-2.0 & Software Development::Debuggers & Other & Fast-Validation \\
pypa/hatch & MIT & Software Development::Build Tools & Build Tools & Fast-Validation \\
pyro-ppl/pyro & Apache-2.0 & Scientific/Engineering::Artificial Intelligence & AI & Fast-Validation \\
python-hyper/h2 & MIT & Internet::WWW/HTTP & Internet & Fast-Validation \\
roboflow/supervision & MIT & Scientific/Engineering::Image Processing & Other & Fast-Validation \\
rytilahti/python-miio & GPL-3.0 & Home Automation & Other & Fast-Validation \\
saleweaver/python-amazon-sp-api & MIT & Internet::WWW/HTTP & Internet & Fast-Validation \\
scrapy/scrapy & BSD-3-Clause & Software Development::Libraries & Libraries & Fast-Validation \\
scverse/scanpy & BSD-3-Clause & Scientific/Engineering::Bio-Informatics & Other & Fast-Validation \\
slackapi/bolt-python & MIT & Communications::Chat & Other & Fast-Validation \\
slackapi/python-slack-sdk & MIT & Communications::Chat & Other & Fast-Validation \\
snowflakedb/snowflake-connector-python & Apache-2.0 & Software Development::Libraries & Libraries & Fast-Validation \\
softlayer/softlayer-python & MIT & Software Development::Libraries & Libraries & Fast-Validation \\
spec-first/connexion & Apache-2.0 & Internet::WWW/HTTP & Internet & Fast-Validation \\
statsmodels/statsmodels & BSD-3-Clause & Scientific/Engineering::Information Analysis & Other & Fast-Validation \\
tfranzel/drf-spectacular & BSD-3-Clause & Software Development::Documentation & Other & Fast-Validation \\
tobymao/sqlglot & MIT & Database::Database Engines/Servers & Database & Fast-Validation \\
tornadoweb/tornado & Apache-2.0 & Internet::WWW/HTTP & Internet & Fast-Validation \\
tortoise/tortoise-orm & Apache-2.0 & Database::Front-Ends & Database & Fast-Validation \\
wagtail/wagtail & BSD-3-Clause & Internet::WWW/HTTP & Internet & Fast-Validation \\ \hline
\end{tabular}
}
\caption{
The information of the repositories involved in FEABench.
}
\label{tab:repo-info}
\end{table*}

%% file: table/repo_filter_stat.tex
\begin{table*}[ht]
\centering
\resizebox{0.8\textwidth}{!}{
\begin{tabular}{l|l|rr|rr|rr}
\hline
\multicolumn{1}{c}{\textbf{Repository Name}} & \multicolumn{1}{c}{\textbf{Class (Short)}} & \multicolumn{1}{c}{\textbf{Max PR No.}} & \multicolumn{1}{c}{\textbf{\# PR}} & \multicolumn{1}{c}{\textbf{\# All tasks}} & \multicolumn{1}{c}{\textbf{\# Candidates}} & \multicolumn{1}{c}{\textbf{\# in Full}} & \multicolumn{1}{c}{\textbf{\# in Lite}} \\ \hline
sympy/sympy & Mathematics & 27454 & 12857 & 6034 & 819 & 239 & 46 \\
joke2k/faker & Testing & 2144 & 1386 & 518 & 186 & 126 & 8 \\
conan-io/conan & Build Tools & 17534 & 6185 & 3715 & 473 & 124 & 23 \\
tobymao/sqlglot & Database & 4597 & 2463 & 1996 & 185 & 116 & 14 \\
scikit-learn/scikit-learn & AI & 30289 & 17489 & 4174 & 186 & 83 & 9 \\
pvlib/pvlib-python & Physics & 2336 & 1072 & 444 & 102 & 52 & 10 \\
deepset-ai/haystack & AI & 8669 & 4132 & 1438 & 317 & 49 & 8 \\
Project-MONAI/MONAI & Medical & 8275 & 3468 & 1809 & 463 & 37 & 1 \\
matplotlib/matplotlib & Other & 29140 & 18163 & 4057 & 164 & 34 & 7 \\
sphinx-doc/sphinx & Other & 13196 & 5798 & 1597 & 126 & 30 & 4 \\
googleapis/python-aiplatform & AI & 4830 & 2927 & 826 & 241 & 29 & 1 \\
astropy/astropy & Physics & 17525 & 11205 & 4292 & 405 & 27 & 8 \\
Textualize/textual & Other & 5444 & 2205 & 819 & 139 & 27 & 1 \\
falconry/falcon & Internet & 2425 & 1279 & 499 & 67 & 27 & 3 \\
softlayer/softlayer-python & Libraries & 2207 & 1277 & 661 & 105 & 26 & 2 \\
Textualize/rich & Libraries & 3548 & 1142 & 283 & 41 & 24 & 4 \\
rytilahti/python-miio & Other & 1993 & 979 & 249 & 46 & 23 & 1 \\
sqlfluff/sqlfluff & Other & 6534 & 3448 & 2079 & 296 & 19 & 2 \\
google-deepmind/optax & AI & 1164 & 805 & 174 & 54 & 19 & 3 \\
pydata/xarray & Other & 9879 & 4310 & 1838 & 234 & 17 & 2 \\
boto/boto3 & Libraries & 4371 & 757 & 130 & 38 & 17 & 2 \\
roboflow/supervision & Other & 1773 & 1021 & 94 & 30 & 14 & 7 \\
RDFLib/rdflib & Libraries & 3025 & 1562 & 370 & 21 & 13 & 2 \\
huggingface/datasets & AI & 7342 & 4118 & 792 & 99 & 12 & 2 \\
aws-cloudformation/cfn-lint & Other & 3898 & 2583 & 790 & 92 & 11 & 0 \\
boto/botocore & Libraries & 3331 & 2145 & 777 & 83 & 11 & 0 \\
pgmpy/pgmpy & AI & 1887 & 902 & 368 & 76 & 11 & 2 \\
huggingface/huggingface\_hub & AI & 2683 & 1559 & 586 & 130 & 10 & 1 \\
prometheus/client\_python & Other & 1080 & 429 & 134 & 14 & 10 & 1 \\
pypa/hatch & Build Tools & 1860 & 721 & 313 & 34 & 9 & 1 \\
scrapy/scrapy & Libraries & 6598 & 3253 & 846 & 53 & 9 & 2 \\
slackapi/python-slack-sdk & Other & 1627 & 764 & 266 & 29 & 9 & 0 \\
django/django & Internet & 18807 & 18482 & 725 & 32 & 7 & 1 \\
pydicom/pydicom & Medical & 2195 & 979 & 494 & 42 & 7 & 1 \\
pylint-dev/pylint & Other & 10168 & 4431 & 1859 & 79 & 7 & 1 \\
embeddings-benchmark/mteb & AI & 1730 & 1050 & 137 & 15 & 7 & 1 \\
python-hyper/h2 & Internet & 1291 & 1072 & 144 & 11 & 6 & 0 \\
mwaskom/seaborn & Other & 3798 & 1093 & 469 & 40 & 5 & 0 \\
pyvista/pyvista & Other & 7045 & 3837 & 1149 & 292 & 5 & 2 \\
dpkp/kafka-python & Libraries & 2442 & 937 & 212 & 17 & 5 & 0 \\
lark-parser/lark & Other & 1503 & 467 & 135 & 10 & 5 & 2 \\
astronomer/astronomer-cosmos & Build Tools & 1439 & 754 & 270 & 39 & 4 & 1 \\
fairlearn/fairlearn & AI & 1472 & 937 & 229 & 29 & 4 & 1 \\
huggingface/accelerate & AI & 3293 & 1607 & 352 & 67 & 4 & 2 \\
docker/docker-py & Libraries & 3297 & 1554 & 538 & 24 & 3 & 1 \\
huggingface/trl & AI & 2550 & 1181 & 248 & 29 & 3 & 0 \\
joblib/joblib & Libraries & 1641 & 701 & 250 & 11 & 3 & 0 \\
open-mmlab/mmengine & Other & 1620 & 1013 & 350 & 59 & 3 & 1 \\
openvinotoolkit/datumaro & Other & 1689 & 1388 & 521 & 98 & 3 & 0 \\
pygments/pygments & Other & 2837 & 886 & 346 & 79 & 3 & 1 \\
pyocd/pyOCD & Other & 1734 & 954 & 175 & 11 & 3 & 0 \\
pyro-ppl/pyro & AI & 3413 & 2302 & 1038 & 369 & 3 & 0 \\
tortoise/tortoise-orm & Database & 1840 & 607 & 283 & 46 & 3 & 0 \\
DLR-RM/stable-baselines3 & AI & 2069 & 534 & 199 & 23 & 2 & 0 \\
EleutherAI/lm-evaluation-harness & AI & 2609 & 961 & 52 & 9 & 2 & 1 \\
PyThaiNLP/pythainlp & Other & 1056 & 667 & 269 & 39 & 2 & 1 \\
TileDB-Inc/TileDB-Py & Libraries & 2128 & 1184 & 488 & 34 & 2 & 0 \\
atlassian-api/atlassian-python-api & Internet & 1476 & 849 & 49 & 14 & 2 & 0 \\
aws/sagemaker-python-sdk & AI & 4987 & 3219 & 715 & 116 & 2 & 2 \\
googleapis/python-bigquery & Internet & 2102 & 1372 & 358 & 65 & 2 & 0 \\
gradio-app/gradio & Other & 10271 & 4707 & 1210 & 81 & 2 & 0 \\
graphql-python/graphene & Libraries & 1583 & 68 & 19 & 2 & 2 & 1 \\
prompt-toolkit/python-prompt-toolkit & Other & 1949 & 713 & 49 & 3 & 2 & 0 \\
snowflakedb/snowflake-connector-python & Libraries & 2127 & 1311 & 466 & 22 & 2 & 0 \\
spec-first/connexion & Internet & 2011 & 879 & 327 & 31 & 2 & 0 \\
statsmodels/statsmodels & Other & 9462 & 3793 & 1444 & 121 & 2 & 0 \\
tornadoweb/tornado & Internet & 3452 & 1573 & 341 & 16 & 2 & 0 \\
pallets/flask & Internet & 5640 & 2604 & 404 & 17 & 1 & 0 \\
pylint-dev/astroid & Libraries & 2669 & 1810 & 301 & 8 & 1 & 0 \\
Aider-AI/aider & Other & 2767 & 346 & 31 & 1 & 1 & 0 \\
Cog-Creators/Red-DiscordBot & Other & 6499 & 4060 & 157 & 18 & 1 & 0 \\
aws-powertools/powertools-lambda-python & Libraries & 5814 & 4398 & 517 & 115 & 1 & 1 \\
biopragmatics/bioregistry & Other & 1346 & 784 & 156 & 28 & 1 & 0 \\
cocotb/cocotb & Other & 4345 & 2137 & 493 & 34 & 1 & 0 \\
facebookresearch/hydra & Libraries & 3005 & 1301 & 442 & 35 & 1 & 0 \\
huggingface/pytorch-image-models & Libraries & 2398 & 551 & 85 & 17 & 1 & 1 \\
minio/minio-py & Libraries & 1472 & 890 & 242 & 14 & 1 & 0 \\
pre-commit/pre-commit & Other & 3382 & 1222 & 527 & 23 & 1 & 0 \\
saleweaver/python-amazon-sp-api & Internet & 1638 & 1027 & 21 & 9 & 1 & 0 \\
scverse/scanpy & Other & 3427 & 1535 & 516 & 35 & 1 & 0 \\
slackapi/bolt-python & Other & 1234 & 446 & 159 & 9 & 1 & 0 \\
tfranzel/drf-spectacular & Other & 1362 & 345 & 135 & 12 & 1 & 1 \\
wagtail/wagtail & Internet & 12732 & 7070 & 1531 & 213 & 1 & 0 \\ \hline
\end{tabular}
}
\caption{
Statistics of how many PRs (task instances) are left during the data collection procedures.
}
\label{tab:repo-filter-stat}
\end{table*}